# Silicon-Nitride Platform for Narrowband Entangled Photon Generation

Sven Ramelow[1,2,*,†], Alessandro Farsi[1,*,†], Stéphane Clemmen[1], Daniel Orquiza[3,5], Kevin Luke[3], Michal Lipson[3,4], Alexander L. Gaeta[1,4]

[1]School of Applied and Engineering Physics, Cornell University, Ithaca, New York 14853, USA
[2]Faculty of Physics, University of Vienna, 1090 Vienna, Austria
[3]School of Electrical and Computer Engineering, Cornell University, Ithaca, New York 14853, USA
[4]Kavli Institute at Cornell for Nanoscale Science, Cornell University, Ithaca, NY 14853
[5] current address: São Paulo State University (UNESP), São João da Boa Vista, Brazil
*corresponding author: sven.ramelow@univie.ac.at
† these authors contributed equally to this work,

**CMOS-compatible photonic chips are highly desirable for real-world quantum optics devices due to their scalability, robustness, and integration with electronics. Despite impressive advances using Silicon nanostructures, challenges remain in reducing their linear and nonlinear losses and in creating narrowband photons necessary for interfacing with quantum memories. Here we demonstrate the potential of the silicon nitride ($Si_3N_4$) platform by realizing an ultracompact, bright, entangled photon-pair source with selectable photon bandwidths down to 30 MHz, which is unprecedented for an integrated source. Leveraging $Si_3N_4$'s moderate thermal expansion, simple temperature control of the chip enables precise wavelength stabilization and tunability without active control. Single-mode photon pairs at 1550 nm are generated at rates exceeding $10^7$ $s^{-1}$ with mW's of pump power and are used to produce time-bin entanglement. Moreover, $Si_3N_4$ allows for operation from the visible to the mid-IR, which make it highly promising for a wide range of integrated quantum photonics applications.**

Quantum optics technology promises disruptive applications in a wide range of fields such as communication, sensing and computing. Fully integrated quantum optical devices [1,2] have recently attracted significant interest due to their promise for scalability, robustness and joint integration with single-photon detectors [3,4] and electronics circuitry. In addition to photon-pair sources on platforms using III-V semiconductors [5,6] or high index glass [7], much of the activity has concentrated on the Silicon (Si) photonics platform [8-14]. However, despite the mature Si-photonics technology and impressive progress for Si microresonator-based sources, the linear and nonlinear losses of Si remain a difficult challenge and hinder, for example the implementation of high-Q microresonators for production of narrowband photons required for interfacing with quantum memories operating in the near infrared and visible regimes. Narrowband photon sources are an important technology for future secure quantum communication networks based on quantum memories [15,16] and become increasingly relevant also for fundamental experiments in single-photon cavity quantum-optomechanics [17]. For quantum memories typically narrow bandwidths well below 100 MHz are necessary defined by the atomic transitions used in their implementation, and even narrower linewidths are typically required in cavity quantum-optomechanics. Standard approaches to achieving such narrow photon bandwidths use either cavity-enhanced spontaneous parametric downconversion (SPDC) in a nonlinear crystal within a stabilized, bulk optical cavity [18-26] or sources based on atomic systems [27-34]. However, due to their size, complexity, lack of wavelength tunablity and the requirement of active cavity or laser locking, these types of sources are unsuitable for future large-scale networks requiring many identical such sources working simultaneously. While some of these drawbacks have recently been remedied with more compact implementations [35,36], platforms based on ultra-compact, fully CMOS-integrated ring-resonators feature distinct advantages: the microresonators and the bus waveguides can be monolithically integrated on the chip, which greatly reduces the size and enables integration of multiple functional components and minimizies sensitivity to environmental perturbations. In addition, since integrated microresonators typically support only a few discrete transverse mode families with widely spaced resonances that can be easily separated, single-mode operation can be readily achieved. Moreover, the advanced dispersion engineering available in integrated platforms enables highly flexible phase-matching. Consequently, fully chip-integrated implementations based on microcavity-enhanced spontaneous four-wave mixing (SFWM) [36], like the one presented here (see Fig. 1), have gained extensive interest [7-14] as a bright, highly compact photon-pair source with a flexible, robust design and the possibility to fabricate many such sources on a single chip. It is possible to include pump filtering [12] on the chip and to generate time-bin entanglement [13]. Moreover, by matching the pump-pulse spectrum to the cavity-linewidth, it is straight-forward to generate pairs with uncorrelated (factorable) joint spectral amplitudes [37]. Nevertheless, despite this rapid progress for chip-integrated photon-pair sources, the Q-factors and corresponding cavity linewidths achieved to date are not sufficient for interfacing to narrow-band quantum memories or quantum optomechanical systems. A promising step in this direction is a recently reported source based on the high-index glass micro-resonators with bandwidths as narrow as 110 MHz [7], albeit without demonstrating precise control over the photon's central wavelength or verifying entanglement generation and no CMOS-compatibility.

Here, we utilize a novel platform for integrated pair-generation based on high-$Q$ $Si_3N_4$ microresonators to demonstrate photon-pair generation with, to our knowledge, the narrowest photon bandwidths (30 MHz) achieved in any fully monolithic, chip-based design. We show that the bandwidths can be selected over a wide range between 30 MHz and 150 MHz and observe high brightness exceeding $10^7$ pairs/s. In addition, we precisely stabilize and tune the central wavelengths of the photons with simple temperature control of the chip with a precision < 10 MHz. Lastly, we demonstrate the photons single-mode character with autocorrelation measurements and verify that the generated photon-pairs are inherently time-bin entangled, which is a central resource for any quantum network.

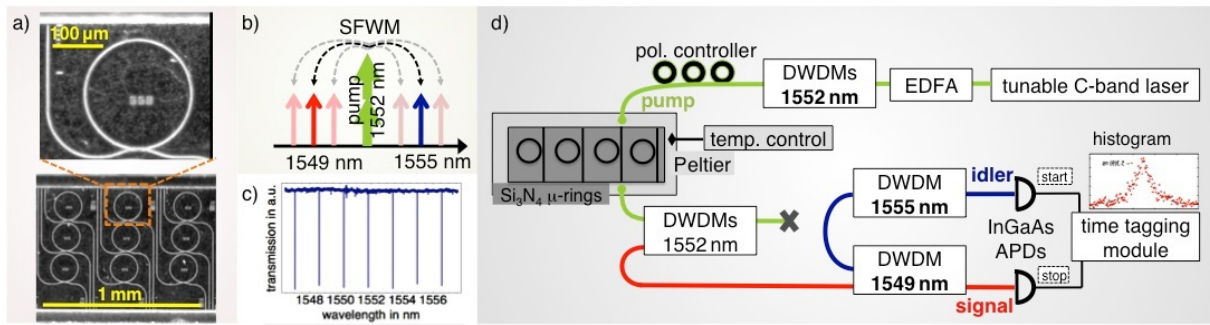

*Figure 1. a) Microscope images of our high-Q $Si_3N_4$ microresonators. The ultra-small footprint easily allows 9 resonators to be integrated onto 1 $mm^2$ of the chip. b) Principle of cavity-enhanced spontaneous four-wave mixing (SFWM) generating photon pairs in adjacent resonances from the central pump wavelength. c) Measured transmission of one of our microresonators for which the high intrinsic Q-factor results in sharp resonances allowing for narrowband photon-pair generation. d) Experimental setup for generating, tuning and characterizing photon pairs with a $Si_3N_4$ chip source.*

Our narrow-band photon-pair source is based on integrated $Si_3N_4$ photonics, which represents a fully CMOS-compatible integrated-optics platform [38] with transparency extending from the 400 nm to 8 $\mu m$ and is highly suited for a wide range of quantum optics applications. Indeed, a number of very recent works have started to demonstrate the unique potential the $Si_3N_4$ platform offers for integrated quantum optics [39-42]. We use low-loss, high-confinement $Si_3N_4$ nanowaveguides with a $SiO_2$ cladding fabricated on standard Si wafers [43]. Such high-confinement waveguides feature moderately high nonlinearities of $\gamma$ = 1 $W^{-1}m^{-1}$ which allows for efficient nonlinear interactions that can be vastly enhanced by microresonators. Due to the ultralow waveguide loss, intrinsic $Q$-factors as high as $Q = 7 \times 10^7$ are used here. All fabrication details are described in [43]. A distinguishing advantage of $Si_3N_4$ is its wide bandgap with no two-photon absorption for wavelengths above 800 nm, which allows for strong pumping without adverse effects, even above the optical parametric oscillation threshold. This is in contrast to Si microresonators in which two-photon-absorption results in maximal pair generation rates saturating at around $10^7$ pairs per second, which corresponds to a modal brightness orders of magnitude below the optimum of ~ 0.1 pairs per temporal mode.

**Experimental Setup**

In our experiments the cross-sections of 690×1100 nm result in a zero-group-velocity dispersion point in the C-band. This ensures phase-matching and frequency matching for efficient pair generation over a broad bandwidth around 1550 nm. Our 115-µm radius microresonators (Fig. 1.a) are evanescently coupled to integrated bus waveguides with the same waveguide cross-section and varying coupling gaps, which lead to selectable loaded Q factors and corresponding adjustable bandwidths. We use lensed fibers to couple into and out of the bus waveguide with inverted tapers for mode-matching [44] and coupling losses of typically 2 -3 dB per facet. The free spectral range (FSR) of our microresonators is 200 GHz. The polarization of the input and output light is controlled and analyzed with standard fiber-based polarization controllers. By monitoring the output power of a tunable laser coupled into the chip, we measure the wavelength-dependent transmission and precisely characterize the resonances (Fig. 1c).

To generate photon pairs by SFWM (Fig. 1b), we pump a resonance at 1551.7 nm with a continuous-wave (CW) tunable external-cavity diode laser. It is amplified with an erbium-doped fiber amplifier (EDFA) and passed through narrow filters based on standard 100-GHz dense wavelength division multiplexing (DWDM) components, which suppresses any undesired noise (e.g., from amplified spontaneous emission of the laser or amplifier) by more than 100 dB. Using spectral filtering at the output also with DWDM-components, we collect the generated photon pairs at two resonances separated by 2 FSRs from the pump resonance at 1548.5 nm for the signal and 1554.9 nm for the idler photons (Fig. 1b). The losses of the filtering and out-coupling from the chip are 3.5 dB and 3 dB, respectively. The generated photons are then detected with home-built single-photon detectors based on InGaAs avalanche photodiodes that are cooled to -40°C and run in gated mode with detection efficiencies of 15% and dark-count levels of $10^{-4}$ per ns gate. The arrival times of the photons are measured and analyzed with a time-tagging module with a nominal resolution of 84 ps. The combined jitter of the two single-photon detectors of 350 ps was determined by an independent measurement and has a negligible effect on the measured, much longer temporal width of the cross-correlation peaks (Fig. 2).

**Selectable Narrow Bandwidths**

An important aspect for efficiently interfacing photon sources with quantum memories or with quantum optomechanical systems is not only to achieve sufficiently narrow bandwidths, but also to precisely match the bandwidth of the target system. Our integrated $Si_3N_4$ platform allows fabrication of many resonators with different coupling gaps and loaded $Q$'s on the same chip. Thus, we can readily select the bandwidth of the generated narrowband photons by tuning to a microresonator with a specific coupling gap between the bus waveguide and the resonator, which is accomplished by a simple translation of the chip relative to the lensed fibers. For narrow-band photons generated in bulk optical cavities, this type of tuning is not readily achievable since it would entail significantly changing the cavities intrinsic losses or the reflectivity of their out-coupling mirrors.

To experimentally verify the generation of our narrowband photon pairs and precisely determine their bandwidths, we measured the normalized cross-correlation of selected microresonators (shown in Fig. 2) using pump powers in the bus waveguide of a few mW. We fit the coincidence peaks to the theoretically expected double exponential [18] $g_{si}(\Delta t) \propto \exp[-|\Delta t|/\tau]$. The resulting time constants $\tau$ are related to the FWHM bi-photon bandwidth $\Delta v = 1/(2\pi\tau)$ [18], which are equal to the resonator bandwidth. As shown in Fig. 2, we observe bandwidths from 150 MHz down to 30 MHz corresponding to biphoton-correlation times ∣ between 1 ns and up to 5 ns. These agree well with the directly measured resonance linewidths of the resonators as determined from fitting the linewidths of the resonance transmission with a Lorentzian (insets in Fig. 2).

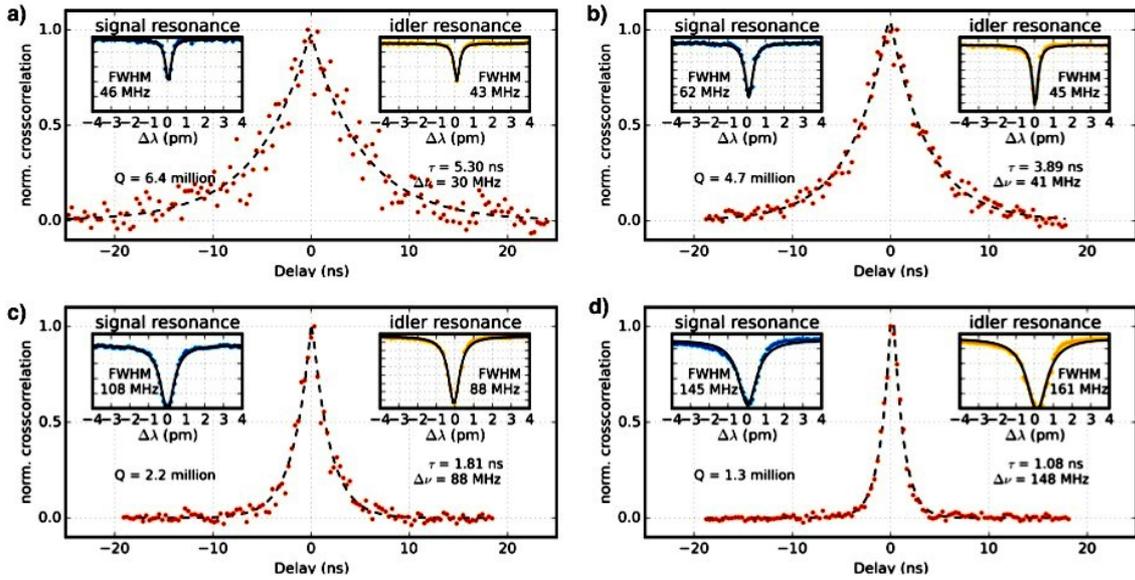

*Figure 2. Bandwidth-selectable photon-pairs in the range between 30 MHz and 150 MHz. (a) – (d) Normalized cross-correlation between signal and idler photons of the generated pairs corresponding to different bandwidths. The data for the coincidence peaks (background corrected and normalized) are fitted with a double exponential, which yields the 1/e coherence time τ and the corresponding bandwidth Δν=1/(2πτ) of the photon pair and loaded Q factor Δν/ν. The insets show transmission scans for the signal (blue) and idler (orange) resonances with Lorentzian fits (black line) and resulting linewidths Δν (FWHM).*

## Wavelength tuning and stabilization

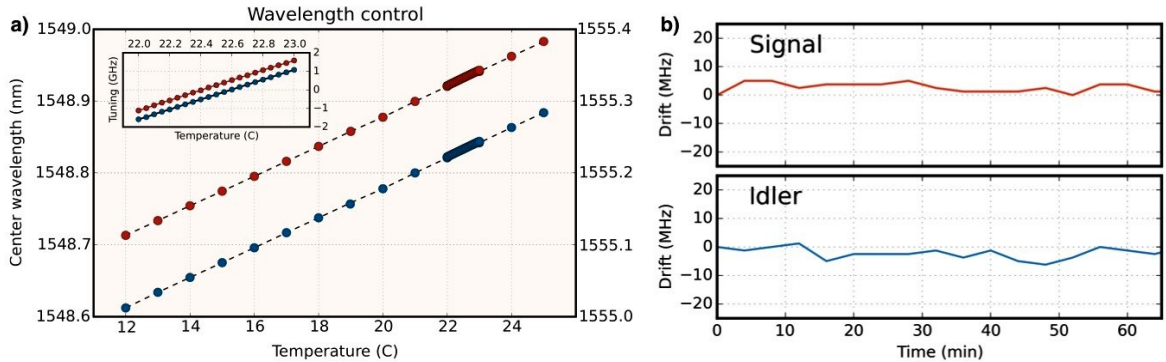

*Figure 3. Wavelength stability and control: measured resonance wavelengths of the signal and idler modes with a tunable laser calibrated by a wavemeter. **a)** Coarse temperature tuning of the signal and idler resonances in steps of 1 K. A range of around 300 pm (35 GHz) can be covered by tuning the temperature by 14 K. **Inset:** Fine-tuning with temperature steps of 0.05 K exemplarily measured between 22 – 23 °C leads to sub-GHz precision in tuning the resonances. **b)** High stability of the signal and idler resonances without locking the resonator by keeping the chip temperature constant while maintaining a pump power of 1 mW (typical for pair generation) coupled to the chip.*

We control and tune the temperature of our chip with a Peltier-element controlled by a standard PID temperature controller with feedback from a 10 kOhm thermistor, which allows precise tuning of the resonance wavelengths of the pump, signal and idler resonances. As shown in Figure 3, both coarse- and fine-tuning of the signal and idler resonances over a broad range of wavelengths are achieved. We measure a tuning coefficient of 21.90 pm/K. Thus a tuning range of 70 K corresponds to a wavelength shift of a full FSR and thereby allows tuning the resonator resonances to any target wavelength in the C-band. Importantly, we can also stabilize the signal and idler resonances of our microresonators by simply maintaining the temperature constant with a PID controller and minimizing airflows, which results in a stability of the signal and idler frequencies to better than ± 10 MHz over more than an hour (Fig. 3b). This stability is achieved with 1 mW of pump power on resonance incident on the ring and illustrates that with minimal temperature control of the chip no external locking of the resonator is necessary, which greatly reduces the overall complexity of our source compared to bulk-cavity enhanced SPDC sources.

**Power dependence, brightness and single-mode operation**

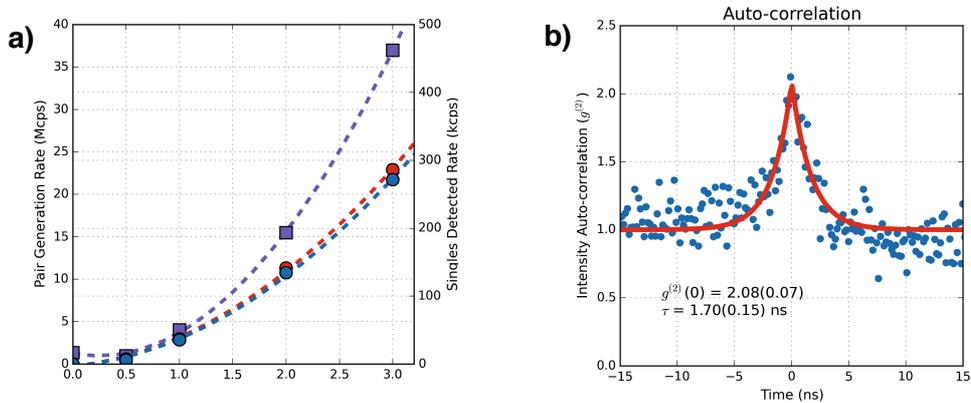

*Figure 4. a) Pump-power-dependent single and coincidence rates show the high brightness of the pair source and the expected quadratic scaling with pump power. b) Measurement of the signal auto-correlation $g_{ss}^{(2)}(t)$ shows the expected peak of $g^{(2)}(0)$ close to the ideal value of 2.*

SFMW in integrated microresonators leads to exceptionally bright pair sources as the theoretically expected pair generation rate $R$ for a given pump power P follows R ∝ $(\gamma P)^2 Q^3 L^{-2}$ [36] with $\gamma$ denoting the effective nonlinearity, $Q$ the quality factor and $L$ the length of the microresonator. Since this rate scales with the 3rd power of the Q-factor, we expect very high brightness in excess of $10^6$ s$^{-1}$ mW$^{-2}$ for our multi-million $Q$ resonators. We confirmed the high brightness and proper power-scaling of our source by measuring the coincidence and singles rates of one of the microresonators for a range of pump-powers coupled to the chip (see Fig. 4). We observe the expected quadratic scaling of the photon rate and, taking into account all losses, infer a pair generation rate of $3.5 \times 10^7$ s$^{-1}$ with 3 mW coupled pump power. Thus, the brightness for this $Q = 2 \times 10^6$ microresonator is $3.9 \times 10^6$ s$^{-1}$mW$^{-2}$, and taking its bandwidth of 90 MHz (FWHM) into account, corresponds to a large spectral brightness of 430,000 s$^{-1}$ mW$^{-2}$ MHz$^{-1}$. An even more useful brightness parameter is the modal brightness (number of pairs generated per temporal mode), which for our system is 0.015 s$^{-1}$ mW$^{-2}$. Thus, to achieve 0.1 pairs per temporal mode as long as $2\tau$, which is an optimal generation rate that maximizes counts but maintains multi-pair emissions at a reasonable level of around 10%, only 2.5 mW of pump power is necessary, which can be readily achieved with diode lasers.

An important aspect of the quality of a photon pair source is its single-mode character and pump-induced noise. A conceptually elegant and stringent way to characterize these properties is to measure the autocorrelation function of the signal photons. One would ideally observe an autocorrelation peak with a $g^{(2)}(0) = 2$. Our measurement (see Fig. 2 b) yields a value very close to this ideal, and thus confirms the single-mode character of our source and negligible pump-induced noise.

## Verification of time-bin entanglement

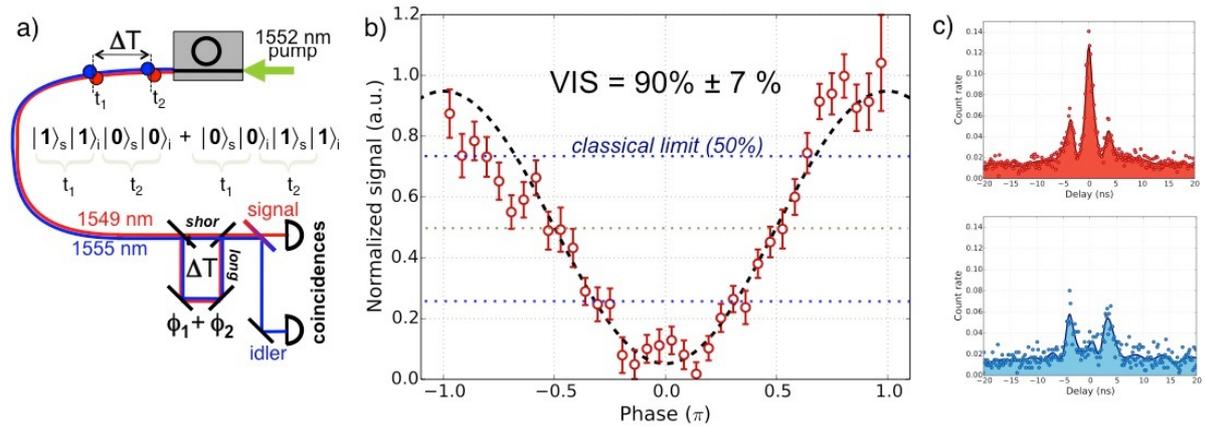

*Figure 5. Verification of time-bin entanglement. a) Principle and setup for verifying time-bin entanglement. b) Measured time-bin entanglement fringe with a sinusoidal fit (dashed line) yielding a high visibility well above the classical and CHSH limit. c) Raw coincidence pattern for the phases of maximum and minimum number of coincidences in the middle peak.*

Entanglement is arguably the most essential ingredient for quantum repeater based quantum networks and for quantum communications, for which time-bin entanglement [45] has been proven to be an excellent encoding. While both polarization and time-bin entanglement can be well transmitted through long fibers, the generation of time-bin entanglement is the most straightforward of all types of photonic entanglement, since it is naturally produced in any parametric pair-production process (e.g. SPDC or SFWM) as a consequence of energy conservation. Its analysis can be achieved for example with post-selection and unbalanced interferometers [45], which leads to three distinct coincidence peaks in the cross-correlation function separated by the unbalanced interferometers' time-delay $\Delta T$ (see Fig. 5). For a time-bin entangled two-photon state of the form $|1\rangle_s|1\rangle_i|0\rangle_s|0\rangle_i + |0\rangle_s|0\rangle_i|1\rangle_s|1\rangle_i$ [the first (second) pair of kets denotes photons in the first (second) time-bin] the number of pairs detected in the middle peak contains all those coincidences where the signal (idler) photon goes the long (short) and short (long) paths, respectively, and will for a time-bin entangled state be sinusoidal dependent on the sum of the phases $\phi_1+\phi_2$ of the unbalanced interferometers. A measurement of this two-photon interference-fringe with a visibility, defined as (maximum – minimum) / (maximum + minimum), above 50% will then verify entanglement [46].

We verify that the generated photon pairs at 1549 nm and 1555 nm are indeed produced in a high-fidelity time-bin entangled state close $|1\rangle_s|1\rangle_i|0\rangle_s|0\rangle_i + |0\rangle_s|0\rangle_i|1\rangle_s|1\rangle$ by wavelength-multiplexing the two unbalanced interferometers for each photon into one fiber-based unbalanced interferometer device (Fig. 5a) and measure the number of coincidences in the middle-peak as a function of the sum of the phase for each photon (Fig. 5b). The phase is controlled by periodically changing the voltage of a piezoelectric ring on which the fiber in the long arm of our interferometer is tightly wrapped attached. After correcting for systematic errors like accidental coincidences, detector dark-counts, and small overall signal variations, the resulting time-bin entanglement fringe has a visibility of 90 ± 7% (Fig. 5b), which clearly verifies the high-fidelity time-bin entangled state produced by our source. We attribute the measured reduction from unit visibility to non-perfect beam-splitter ratios and polarization alignment of our unbalanced fiber interferometer and to residual phase noise of our pump laser.

## Conclusions and outlook

Our results show for the first time the highly promising nature of the $Si_3N_4$ platform for quantum optics applications by implementing a fully monolithic source of unprecedentedly narrow-band photon pairs that are time-bin entangled and can be precisely tuned and stabilized in wavelength. Since the *Q*-factors of our $Si_3N_4$ micro-resonators are not yet material absorption limited [43,47], further refinement of the fabrication techniques promises even higher *Q*'s, which would yield photon bandwidths in the sub-10-MHz regime. Q-factors in high-confinement $Si_3N_4$ in excess of 15 million have indeed already been reported [47]. Moreover, using integrated heaters and temperature sensors embedded in the chip would result in even more robust temperature stability and control, which could be further enhanced by using athermal designs [48]. To further show the promise of the $Si_3N_4$ platform for quantum networks, a straightforward next step is to connect two such sources and demonstrate indistinguishability between 2 heralded photons from fully independent chips. Finally, the broad transparency of $Si_3N_4$ combined with the excellent dispersion engineering possible for $Si_3N_4$ waveguides will allow pair generation not only in the C-Band but anywhere from the visible to the mid-infrared making it a very flexible integrated nonlinear quantum-optics platform to interface to a wide range of quantum applications and devices.